# The qualitative analysis of the impact of media delay on the control of infectious disease

Dongmei Li[1] Yue Wu[2] Panpan Wen[1] Weihua Liu[1]

(1 Applied Science College, Harbin University of Science Technology, Harbin 150080 china

2 Corresponding Author; Schlumberger WesternGeco Geosolution, Houston, Texas, USA)

**Abstract**: In this paper, we consider the impact of time delay by media on the control of the disease. We set up a class of SISM epidemic model with the time delay and the cumulative density of awareness caused by media. The sufficient condition of global asymptotic stability of disease-free equilibrium is approved. We get the global stability of the epidemic equilibrium and the existence conditions of Hopf bifurcation. Numerical simulations are presented to illustrate the analytical results. Finally, we analysis the influence of parameters on the control of infectious disease by combining the data of H1N1. By shortening the time of media lag, increasing transmission rate of media and the implementation rate of the media project, the spread of disease will be controlled effectively.

# 1. Introduction

If a new infectious disease occurs, infected numbers, death numbers, infection disease symptoms and other related information is reported by media, it makes people strengthen the prevention awareness, choose consciously preventive measures and minimize the risk of contact with infection individuals. These precautions reduce the chances of being infected, so the media plays a important role in preventing epidemic[1].

The propaganda of disease mainly affects the effective contact rate of susceptible individuals, it can increase the awareness of prevention and reduce the spread of disease by adjusting the media propaganda[2,3,4]. In recent years, due to the awareness being cumulated by media, it appears that the corresponding research results about controlling disease effectively[5,6]. Under the influence of media information, the susceptible individuals are divided into two groups of “aware” and “unaware”. And the susceptible individuals in the unaware group will eventually all convert into susceptible individuals who have awareness. The cumulative density of awareness is introduced into the model as an independent dynamic variable, SISM model is established and the existence conditions of equilibrium are given, the global stability of equilibrium is proved by structuring Liapunov function[7,19]. For the SISM model with the delay of latency, the results show that the stability changes and the model appears Hopf bifurcation, with the change of the parameters of latency. Considering that a small part of recovered individuals are still unawared susceptible individuals, while the rest restore to awareness susceptible individuals. The global stability of equilibrium and the existence conditions of Hopf bifurcation are studied by establishing SISM model. From the normal form theory, we infer the direction of Hopf bifurcation, the stability and the approximate expression of the periodic solution[9]. When the disease occurs, the media does not report immediately, and it is reported after a certain time. In addition, we do not take timely measures to prevent disease after media reports the infectious disease. Under the situation of different delay parameters, the global asymptotic stability of the equilibrium was proved by structuring Liapunov function in SISM model.

On the basis of existing results, the influence of latency and information time lag is researched. We establishes double delay SISM infectious disease model, and the stability of equilibrium and the existence conditions of Hopf bifurcation are studied in this paper.

## 2. The Model

There exists a latent period for the infectious disease, when the media reports the disease. At time $t$ exposed individuals show symptoms after latent period $\tau$. In the meantime, we take into account of the case that the media do not report the information about the disease at time $t$, and it is reported after a time delay of $h$. In this paper, we obtain the double delay $S_1IS_2M$ infectious disease model.

$$\begin{cases} \dfrac{dS_1(t)}{dt} = c - \beta S_1(t)I(t-\tau) + \xi_0 S_2(t) - \xi S_2(t)M(t-h) - cS_1(t) + (1-p)vI(t) \\ \dfrac{dI(t)}{dt} = \beta S_1(t)I(t-\tau) - (c+v)I(t) \\ \dfrac{dS_2(t)}{dt} = \xi S_2(t)M(t-h) - (\xi_0 + c)S_2(t) + pvI(t) \\ \dfrac{dM(t)}{dt} = \mu I(t) - \mu_0 M(t) \end{cases} \tag{1}$$

where $S_1(t)$, $S_2(t)$ and $I(t)$ represent the number of the unaware susceptible individuals, aware susceptible individuals and infection individuals at time $t$, respectively; $M(t)$ represents that the cumulative density of awareness programs driven by media at time $t$; $c$ represents the rate of constant input and natural death; $\beta$ is the effective contact rate of susceptible individuals with the infective individuals; $\xi_0$ represents the rate of transfer from aware susceptible individuals to unaware; $\xi$ represents the dissemination rate of awareness among susceptible individuals; $\mu$ represents the rate with which media is being implemented; $\mu_0$ represents the depletion rate of that program due to the case of invalid media. All parameters are non-negative.

Let $N(t) = S_1(t) + I(t) + S_2(t)$, summing up the four equations of model (1), we have $\dfrac{dN(t)}{dt} = c - cN(t)$. Then

$$\limsup_{t\to+\infty} N(t) = 1 \tag{2}$$

Considering the limit equation of model (1)

$$\begin{cases} \dfrac{dI(t)}{dt} = \beta(1-S_2(t)-I(t))I(t-\tau)-cI(t)-vI(t) \\ \dfrac{dS_2(t)}{dt} = \xi(1-S_2(t)-I(t))M(t-h)-\xi_0 S_2(t)-cS_2(t)c)S_2(t) \\ \dfrac{dM(t)}{dt} = \mu I(t)-\mu_0 M(t) \end{cases} \tag{3}$$

Let $l=\max(\tau,h)$, supposing that $C$ represents all continuous mapping: $\varphi:[-l,0]\to R^3$ constitutes a Banach space, and denoting norm $\|\varphi\|=\max\{\sup_{-l\le\theta\le 0}|\varphi_1(\theta)|, \sup_{-l\le\theta\le 0}|\varphi_2(\theta)|, \sup_{-l\le\theta\le 0}|\varphi_3(\theta)|\}$, where $\varphi=(\varphi_1,\varphi_2,\varphi_3)$.

According to the mean of biology, it follows that $\varphi=(\varphi_1,\varphi_2,\varphi_3)\in C_+$, $C_+=\{\varphi_i(\theta)>0,\forall\theta\in[-l,0], i=1,2,3\}$.

The initial condition of model (3) is

$$I(\theta)=\varphi_1(\theta), S_2(\theta)=\varphi_2(\theta), M(\theta)=\varphi_3(\theta), -l\le\theta\le 0$$

From the third equation of model (3), we have

$$\frac{dM(t)}{dt}\le\mu-\mu_0 M(t)$$

Combining equation (2) and (4), the feasible region of model (3) is

$$\Omega=\{(I,S_2,M)\,|\,0<I+S_2<1, 0\le M\le\mu/\mu_0\}$$

# 3. The main research results

The basic reproduction number of model (3) is

$$R_0=\frac{\beta}{v+c}$$

**Lemma 1**[11] Considering the following equation

$$x'(t)=ax(t-\tau)-bx(t),$$

Where $a,b,\tau>0$, we have:

(i) if $a<b$, then $\lim_{t\to\infty}x(t)=0$;

(ii) if $a>b$, then $\lim_{t\to\infty}x(t)=+\infty$.

The equilibria of model (3) satisfies the following equations

$$\begin{cases}\beta(1-S_2-I)I-cI-vI=0\\ \xi(1-S_2-I)M-\xi_0S_2-cS_2+pvI=0\\ \mu I-\mu_0M=0\end{cases}$$

We obtain the disease-free equilibrium $E_0(0,0,0)$ and the endemic equilibrium $E^*(I^*,S_2^*,M^*)$, where $I^*=\dfrac{\mu_0(\xi_0+c)(R_0-1)}{\xi\mu+(\xi_0+c+pv)R_0\mu_0}$, $S_2^*=1-1/R_0-I^*$, $M^*=\dfrac{\mu}{\mu_0}I^*$

**Theorem 1** If $R_0\le 1$, then the disease-free equilibrium $E_0(0,0,0)$ of model (3) exists. If $R_0>1$, then the unique endemic equilibrium $E^*(I^*,S_2^*,M^*)$ of model (3) exists.

**Theorem 2** If $R_0\le 1$, then the disease-free equilibrium $E_0$ of model (3) is globally asymptotically stable.

Proof: The linearization equation of model (3) at $E_0$ as follows

$$\begin{cases}\dfrac{dI(t)}{dt}=\beta I(t-\tau)-cI(t)-vI(t)\\ \dfrac{dS_2(t)}{dt}=\xi M(t-h)-\xi_0S_2(t)-cS_2(t)+pvI(t)\\ \dfrac{dM(t)}{dt}=\mu I(t)-\mu_0(t)M(t)\end{cases} \tag{5}$$

The characteristic equation of (5) is

$$(\lambda+\mu_0)(\lambda+\xi_0+c)(\lambda+v+c-\beta e^{-\lambda\tau})=0$$

with characteristic roots $\lambda_1=-\mu_0$, $\lambda_2=-(\xi_0+c)$. Next, we consider equation as follows

$$\lambda+v+c-\beta e^{-\lambda\tau}=0 \tag{6}$$

Suppose that the equation (6) has a complex root with non-negative real part $\lambda=\alpha+i\omega$, i.e. $\alpha>0$. Substituting it into equation (6), and separating real and imaginary parts of equation (6) gives that

$$\begin{cases}\alpha+v+c=\beta e^{-\alpha\tau}\cos\omega\tau\\ \omega=\beta e^{-\alpha\tau}\sin\omega\tau\end{cases} \tag{7}$$

We square both sides of the equation (7) and add squared above equations to obtain the following equation

$$(\alpha+v+c)^2+\omega^2=(\beta e^{-\alpha\tau})^2 \tag{8}$$

It follows from $R_0 \le 1$ that $\beta \le v + c$, so we have $\beta e^{-\alpha\tau} < v + c$, it is in contradiction with equation (8). Then the roots of equation (6) only have negative real part. Consequently, the disease-free equilibrium $E_0$ of model (3) is locally asymptotically stable.

Considering the following Liapunov function:

$$V(t) = I(t)$$

Now differentiating $V$ with respect to $t$ along to the model (3) in the region $\Omega$, we get

$$V'(t) = \beta\left(1 - S_2(t) - I(t)\right) I(t-\tau) - (v+c) I(t) \le \beta I(t-\tau) - (v+c) I(t)$$

According to $R_0 \le 1$ and Lemma 1, it follows that $\lim_{t\to\infty} V(t) = 0$. Clearly, the largest invariant set of model (3) is $\{I = 0\}$. Therefore, the disease-free equilibrium $E_0$ of model (3) is globally asymptotically stable. This completes the proof.

Let $I = I^* + i$, $S_2 = S_2^* + s_2$, $M = M^* + m$, we obtain the linearization system of model (3) at $E^*$

$$\begin{cases} \dfrac{di(t)}{dt} = \beta(1 - S_2^* - I^*) i(t-\tau) - \beta I^* s_2(t) - (c + v + \beta I^*) i(t) \\ \dfrac{ds_2(t)}{dt} = (pv - \xi M^*) i(t) - (\xi_0 + c + \xi M^*) s_2(t) + \xi(1 - S_2^* - I^*) m(t-h) \\ \dfrac{dm(t)}{dt} = \mu i(t) - \mu_0 m(t) \end{cases} \tag{9}$$

The characteristic equation of equation (9) is

$$\begin{aligned} &\lambda^3 + (\mu_0 + \xi M^* + \xi_0 + 2c + \beta I^* + v)\lambda^2 + [\mu_0(\xi M^* + \xi_0 + 2c + \beta I^* + v) \\ &\quad + \beta I^*(\xi_0 + c + pv) + (c+v)(\xi M^* + \xi_0 + c)]\lambda + \mu_0[(c+v)(\xi M^* + \\ &\quad \xi_0 + c) + \beta I^*(\xi_0 + c + pv)] - [\lambda^2 + (\xi M^* + \xi_0 + c + \mu_0)\lambda + \mu_0(\xi M^* \\ &\quad + \xi_0 + c)](v+c)e^{-\lambda\tau} + (v+c)\mu\xi I^* e^{-\lambda h} = 0 \end{aligned} \tag{10}$$

In the following, we will discuss the characteristic roots of the equation (10) and the stability of $E^*$ while $\tau$ and $h$ are different.

Case(I) $\tau = h = 0$

Simplifying the characteristic equation (10), we have

$$\lambda^3 + a_1\lambda^2 + a_2\lambda + a_3 = 0 \tag{11}$$

where $a_1 = \mu_0 + \xi_0 + c + \beta I^* + \xi M^*$, $a_2 = \mu_0(\xi M^* + \xi_0 + c) + \beta I^*(\xi_0 + c + pv + \mu_0)$, $a_3 = \mu_0 \beta I^*(\xi_0 + c + pv) + \mu\xi I^*(v + c)$.

According to Routh-Hurwitz Criterion, we have the following theorem.

**Theorem 3** When $R_0 > 1$ and $\tau = h = 0$, the endemic equilibrium $E^*$ of model (3) is locally asymptotically stable if and only if $a_1 a_2 > a_3$.

Case(II) $\tau = 0$, $h > 0$

Simplifying the characteristic equation (10), we have

$$\begin{aligned}&\lambda^3 + (\mu_0 + \xi M^* + \xi_0 + c + \beta I^*)\lambda^2 + [\mu_0(\xi M^* + \xi_0 + c + \beta I^*) \\ &+ \beta I^*(\xi_0 + c + pv)]\lambda + \mu_0 \beta I^*(\xi_0 + c + pv) + (v + c)\mu\xi I^* e^{-\lambda h} = 0\end{aligned} \tag{12}$$

Suppose that $\lambda = \pm i\omega(h)(\omega > 0)$ are a pair of purely imaginary roots of equation (12). Substituting it into equation (12), and separating real and imaginary parts of equation (12) gives that

$$\begin{cases} -\omega^3 + [\mu_0(\xi M^* + \xi_0 + c + \beta I^*) + \beta I^*(\xi_0 + c + pv) + (c + v)(\xi M^* + \xi_0 \\ \quad + c)]\omega = (c + v)\mu\xi I \sin\omega h \\ (\mu_0 + \xi M^* + \xi_0 + c + \beta I^*)\omega^2 - \mu_0 \beta I^*(\xi_0 + c + pv) = (c + v)\mu\xi I^* \cos\omega h \end{cases} \tag{13}$$

We square both sides of each equation above and add the squared above equations to obtain the following equation

$$f(\omega^2) = (\omega^2)^3 + q_1(\omega^2)^2 + q_2\omega^2 + q_3 = 0 \tag{14}$$

where $q_1 = \mu_0^2 + (\xi M^* + \xi_0 + c + \beta I^*)^2 - 2\beta I^*(\xi_0 + c + pv)$,

$q_2 = \mu_0^2(\xi M^* + \xi_0 + c + \beta I^*)^2 + \beta I^*(\xi_0 + c + pv)[\beta I^*(\xi_0 + c + pv) - 2\mu_0^2]$,

$q_3 = \mu_0^2 \beta^2 I^{*2}(\xi_0 + c + pv)^2 - \mu^2 \xi^2 I^{*2}(v + c)^2$.

It is knowing that $f(0) = q_3$ and $\lim_{\omega^2 \to +\infty} f(\omega^2) = +\infty$.

When $q_3 < 0$, at least equation (14) exists a positive root.

When $q_3 > 0$, we take derivative to the both sides of the equation (14) about $\omega^2$, so

$$f'(\omega^2) = 3(\omega^2)^2 + 2q_1(\omega^2) + q_2 \tag{15}$$

Let $\Delta$ be discriminant, where $\Delta = 4(q_1^2 - 3q_2)$.

When $\Delta < 0$, equation (15) has no real root, then equation (14) has no positive real root.

When $\Delta \geq 0$, if $q_1 > 0, q_2 > 0$, equation (15) has no positive real root, then equation (14) has no positive real root; if $q_1$ and $q_2$ are the other cases, equation (15) has at least a positive real root, i.e. $f'(\varpi^2) = 0$, and $f(\varpi^2) \leq 0$, then equation (14) has at least a positive root.

Suppose that equation (14) has finite positive roots $\omega_k^2 \in [0,\bar{\omega}^2]$ $(k=1,2,\cdots,m)$ in the interval $[0,\bar{\omega}^2]$, substituting it into the second formula of equation (12). We obtain

$$h_k^j = \frac{1}{\omega_k}\arccos\left\{\frac{(\mu_0+\xi_0+c+\xi M^*+\beta I^*)\omega_k^2}{(v+c)\mu\xi I^*} - \frac{\mu_0\beta(\xi_0+c+pv)}{(v+c)\mu\xi}\right\} + \frac{2j\pi}{\omega_k}$$
$$(k=1,2,3,...m;\ j=0,1,2,...)$$

Let $h_k^j$ satisfy $\alpha(h_k^j)=0,\ \omega(h_k^j)=\omega_k$ and $h_1^0=\min\{h_k^j\}$.

Then, we have

**Theorem 4** If $R_0>1$, $\tau=0$, $h>0$ hold, we have

1) When $q_3>0$, $\Delta<0$ or $\Delta\ge 0$ and $q_i>0(i=1,2)$, for all $h\ge 0$, the endemic equilibrium $E^*$ of model (3) is asymptotically stable;

2) When $q_3<0$ or $\Delta\ge 0$, $q_i(i=1,2)$ are non-positive at the same time, the limit point of $f(\omega^2)$ is $\bar{\omega}^2$, and when $f(\omega^2)\le 0$, for each $h\in[0,h_1^0)$, the endemic equilibrium $E^*$ of model (3) is asymptotically stable; when $h>h_1^0$, the characteristic roots of equation (12) have positive real part, then the endemic equilibrium $E^*$ of model (3) is unstable;

3) If the condition 2) holds, when $h=h_k^j(k=1,2,3,...,m;\ j\in N)$, then there exists Hopf bifurcation around the endemic equilibrium $E^*$ of the model (3).

Proof 1) According to the roots existence conditions of cubic polynomial, when $q_3>0$, $\Delta<0$ or $\Delta\ge 0$ and $q_i>0(i=1,2)$, then equation (14) has no positive root . By the Cook Theorem[13] we know that all characteristic roots of equation (12) have negative real parts, that is to say, $E^*$ is asymptotically stable, conclusion 1) holds.

2) Suppose that $\lambda(h)=\alpha(h)+i\omega(h)$ is positive root of equation (12). When $\Delta\ge 0$, $q_1$ and $q_2$ are non-positive at the same time, $f'(\omega^2)$ has a positive root $\bar{\omega}^2$, and $f(\omega^2)\le 0$, it follows that equation (14) exists positive roots $\omega_k^2(k=1,2,\cdots,m)$ in $[0,\bar{\omega}]$, let $\omega_0$ satisfy $\alpha(h_1^0)=0,\ \omega(h_1^0)=\omega_0$. According to the continuous dependence of characteristic root $\alpha(h)$ to $h$ and the choice of $h_1^0$, it follows that the critical value $h_1^0$ is $h$ which makes the characteristic root firstly through to imaginary axis. According to the Cook Lemma[13], for each $h\in[0,h_1^0)$, it follows that all the roots of equation (12) have strict negative real part, so $E^*$ is asymptotically stable. When $h>h_1^0$, then the characteristic roots of equation (12) exist positive real part, the endemic equilibrium $E^*$ of model (3) is unstable, conclusion 2) holds.

3) If condition 2) holds, now differentiating equation (12) with respect to $h$, we get

$$\{3\lambda^2+2[\mu_0+\xi M^*+\xi_0+c+\beta I^*]\lambda+[\mu_0(\xi M^*+\xi_0$$
$$+2c+\beta I^*+v)+\beta I^*(\xi_0+c+pv)-\mu_0(v+c)\}(d\lambda/dh)=(v+c)\lambda\mu\xi I^* e^{-\lambda h}$$

Solving the above equation, we have

$$\begin{aligned}\left(\frac{d\lambda}{dh}\right)^{-1} &= \frac{e^{\lambda h}[3\lambda+2(\mu_0+\xi M^*+\xi_0+c+\beta I^*)]}{\mu\xi I^*(v+c)} \\ &+\frac{e^{\lambda h}[\mu_0(\xi M^*+\xi_0+2c+\beta I^*+v)+\beta I^*(\xi_0+c+pv)]}{\lambda\mu\xi I^*(v+c)} \\ &-\frac{(\xi_0+c+\mu_0+\xi M^*)e^{\lambda h}}{\lambda\mu\xi I^*}+\frac{(\xi_0+c+\xi M^*)e^{\lambda h}}{\lambda\mu\xi I^*}-\frac{h}{\lambda}\end{aligned} \tag{16}$$

Substituting $\lambda=i\omega_k$ into equation (16) and combining equation (13), we get

$$\begin{aligned}\left(\frac{d\operatorname{Re}\lambda(h_k^j)}{dh}\right)^{-1}_{h=h_k^j} &= -\frac{\xi M^*+\xi_0+c}{\omega_k\mu\xi I^*}\sin(\omega_k h)+\frac{\xi_0+c+\mu_0+\xi M^*}{\omega_k\mu\xi I^*}\sin(\omega_k h) \\ &+\frac{2[\mu_0+\xi M^*+\xi_0+c+\beta I^*]\cos(\omega_k h)+3\omega_k\sin(\omega_k h)}{\mu\xi I^*(v+c)} \\ &-\frac{\mu_0(\xi M^*+\xi_0+2c+v+\beta I^*)+\beta I^*(\xi_0+c+pv)}{\omega_k\mu\xi I^*(v+c)}\sin(\omega_k h) \\ &=\frac{f'(\omega_k^2)}{[\mu\xi I^*(v+c)]^2}\end{aligned}$$

Due to the choice of $\omega_k$ and the property of quadratic function, we have $f'(\omega_k^2)\neq 0$. Combining transversality condition[13], $sign\left(d\operatorname{Re}\lambda(h_k^j)/dh\right)^{-1}\neq 0$ holds. According to Hopf bifurcation theorem[14], it follows that Hopf bifurcation appears if $h$ through to $h_k^j$, conclusion 3) holds. This completes the proof of theorem 4.

Case(III) $\tau>0,\ h=0$

Simplifying the characteristic equation (10), we have

$$\begin{aligned}&\lambda^3+(\mu_0+\xi M^*+\xi_0+2c+\beta I^*+v)\lambda^2+[\mu_0(\xi M^*+\xi_0+2c+\beta I^*+v)+ \\ &\beta I^*(\xi_0+c+pv)+(v+c)(\xi M^*+\xi_0+c)]\lambda+\mu_0[(v+c)(\xi M^*+\xi_0+c)+ \\ &\beta I^*(\xi_0+c+pv)]+\mu\xi I^*(v+c)-(v+c)[\lambda^2+(\xi M^*+\xi_0+c+\mu_0)\lambda+\mu_0 c \\ &\mu_0(\xi M^*+\xi_0)]e^{-\lambda\tau}=0\end{aligned} \tag{17}$$

Suppose that $\lambda=\pm i\omega(h)(\omega>0)$ are a pair of purely imaginary roots of equation (17). Substituting it into equation (17), and separating real and imaginary parts, we have

$$f(\omega^2)=(\omega^2)^3+p_1(\omega^2)^2+p_2\omega^2+p_3=0 \tag{18}$$

where $p_1=\mu_0^2+(\xi M^*+\xi_0+c+\beta I^*)^2+2\beta I^*(v-\xi_0-pv)$

$$\begin{aligned}p_2=&\mu_0^2(\xi M^*+\xi_0+c+\beta I^*)^2+2\beta I^*\mu_0^2(c+v)+(\beta I^*)^2(\xi_0+c+pv)^2\\&+2\beta I^*(c+v)(\xi M^*+\xi_0+c)(\xi_0+c+pv)-2\mu_0^2\beta I^*(c+pv+\xi_0)\\&-2(v+c)\mu\xi(\xi M^*+\xi_0+2c+v+\mu_0)\end{aligned}$$

$$\begin{aligned}p_3=&(v+c)^2\mu^2\xi^2I^{*2}+2(v+c)\mu\xi I^*\mu_0[(c+v)(\xi M^*+\xi_0+c)+\beta I^*(\xi_0+c+pv)]\\&+\mu_0^2\beta^2I^{*2}(\xi_0+c+pv)^2+2\mu_0^2\beta I^*(c+v)(\xi M^*+\xi_0+c)(\xi_0+c+pv)\end{aligned}$$

When $R_0>1$, then $p_3>0$. Now differentiating equation (18) with respect to $\omega^2$, we have

$$f'(\omega^2)=3(\omega^2)^2+2p_1(\omega^2)+p_2 \tag{19}$$

Let $\Delta=4(p_1^2-3p_2)$ be the discriminant of equation (19).

Suppose that $\bar{\omega}^2$ is positive root of equation (19), i.e. $f'(\bar{\omega}^2)=0$, and $f(\bar{\omega}^2)<0$, then equation (18) have positive roots $\omega_l^2(l=1,2,\cdots,n)$ in $[0,\bar{\omega}^2]$, we obtain

$$\tau_l^i=\frac{1}{\omega_l}\arctan\left\{\frac{m_2m_3+m_1m_4}{m_1m_3-m_2m_4}\right\}+\frac{ik\pi}{\omega_l},(i=0,1,2,...;l=1,2,3,...,n)$$

where

$$\begin{aligned}m_1=&-\omega_l^3+[\mu_0(\xi M^*+\beta I^*+\xi_0+2c+v)+\beta I^*(\xi_0+c+pv)\\&+(v+c)(\xi M^*+\xi_0+c)]\omega_l\end{aligned}$$

$$\begin{aligned}m_2=&-(\mu_0+\xi M^*+\xi_0+2c+v+\beta I^*)\omega_l^2+\mu_0[(v+c)(\xi M^*+\xi_0+c)\\&+\beta I^*(\xi_0+c+pv)]\end{aligned}$$

$$m_3=\omega_l(v+c)(\xi M^*+\xi_0+c+\mu_0)$$

$$m_4=[\omega_l^2-\mu_0(\xi M^*+\xi_0+c)](v+c)$$

Let $\tau_l^i$ satisfy $\alpha(\tau_l^i)=0,\ \omega(\tau_l^i)=\omega_l$, $\tau_1^0=\min\{\tau_l^i\}$ .we have the following theorem.

**Theorem 5** If $R_0>1$, $\tau>0$, $h=0$ holds, we have

1) When $\Delta<0$ or $\Delta\ge 0$ and $p_1,p_2>0$, for all $\tau\ge 0$, the endemic equilibrium $E^*$ of model (3) is asymptotically stable;

2) When $\Delta \geq 0$, $p_i(i=1,2)$ are not positive at the same time, the limit point of $f(\omega^2)$ is $\bar{\omega}^2$, and when $f(\omega^2) \leq 0$, for each $\tau \in [0, \tau_1^0)$, the endemic equilibrium $E^*$ of model (3) is asymptotically stable; when $\tau > \tau_1^0$, the characteristic roots of equation (18) have positive real part, then the endemic equilibrium $E^*$ is unstable;

3) If the condition 2) and

$$2(\mu_0+\xi M^*+\xi_0+2c+\beta I^*+v)[2m_1m_3m_4+m_2(m_3^2-m_4^2)]\omega_0+[\mu_0(\xi M^*+\xi_0+2c+\beta I^*+v)-3\omega_0^2+(\xi M^*+\xi_0+c)(c+v)+\beta I^*(\xi_0+c+pv)][m_1(m_3^2-m_4^2)-2m_2m_3m_4]-[2\omega_0 m_4+m_3(\xi M^*+\xi_0+c+\mu_0)](c+v)^2(m_3^2+m_4^2)\neq 0$$

4) holds, when $\tau=\tau_l^i(l=1,2,3,...,n;\ i\in N)$, then there exists Hopf bifurcation around the endemic equilibrium $E^*$ of the model (3).

Proof The proof of this theorem can be complete by the method analogous to that used above proof.

3) Now differentiating equation (17) with respect to $\tau$, we get

$$\left(\frac{d\lambda}{d\tau}\right)^{-1}=-\frac{e^{\lambda\tau}[3\lambda+2(\mu_0+\xi M^*+\xi_0+2c+\beta I^*+v)]}{(v+c)[\lambda^2+(\xi M^*+\xi_0+c+\mu_0)\lambda+\mu_0(\xi M^*+\xi_0+c)]}-\frac{\mu_0(\xi M^*+\xi_0+2c+\beta I^*+v)e^{\lambda\tau}}{(v+c)\lambda[\lambda^2+(\xi M^*+\xi_0+c+\mu_0)\lambda+\mu_0(\xi M^*+\xi_0+c)]}-\frac{e^{\lambda\tau}[(\xi M^*+\xi_0+c)(v+c)+\beta I^*(\xi_0+c+pv)]}{(v+c)\lambda[\lambda^2+(\xi M^*+\xi_0+c+\mu_0)\lambda+\mu_0(\xi M^*+\xi_0+c)]}+\frac{2\lambda+\xi M^*+\xi_0+c+\mu_0}{\lambda[\lambda^2+(\xi M^*+\xi_0+c+\mu_0)\lambda+\mu_0(\xi M^*+\xi_0+c)]}-\frac{\tau}{\lambda} \tag{20}$$

Substituting $\lambda = i\omega_l$ into equation (20) , we have

$$\left(\frac{d\,\mathrm{Re}\,\lambda(\tau_l^i)}{d\tau}\right)^{-1}_{\tau=\tau_l^i} = \frac{2(\mu_0+\xi M^*+\xi_0+2c+\beta I^*+v)(m_3\sin\omega_l\tau+m_4\cos\omega_l\tau)}{(v+c)(m_3^2+m_4^2)}$$

$$+\frac{[\mu_0(\xi M^*+\xi_0+2c+\beta I^*+v)-3\omega_l^2](m_3\cos\omega_l\tau-m_4\sin\omega_l\tau)}{\omega_l(v+c)(m_4^2+m_3^2)}$$

$$+\frac{[(\xi M^*+\xi_0+c)(c+v)+\beta I^*(\xi_0+c+pv)](m_3\cos\omega_l\tau-m_4\sin\omega_l\tau)}{\omega_l(v+c)(m_3^2+m_4^2)}$$

$$-\frac{2\omega_l m_4+m_3(\xi M^*+\xi_0+c+\mu_0)}{\omega_l(m_3^2+m_4^2)}$$

$$=\frac{2(\mu_0+\xi M^*+\xi_0+2c+\beta I^*+v)[2m_1m_3m_4+m_2(m_3^2-m_4^2)]}{(v+c)^2(m_3^2+m_4^2)^2}$$

$$+\frac{[\mu_0(\xi M^*+\xi_0+2c+\beta I^*+v)-3\omega_l^2][m_1(m_3^2-m_4^2)-2m_2m_3m_4]}{\omega_l(v+c)^2(m_4^2+m_3^2)^2}$$

$$+\frac{[(\xi M^*+\xi_0+c)(c+v)+\beta I^*(\xi_0+c+pv)][m_1(m_3^2-m_4^2)-2m_2m_3m_4]}{\omega_l(v+c)^2(m_3^2+m_4^2)^2}$$

$$-\frac{2\omega_l m_4+m_3(\xi M^*+\xi_0+c+\mu_0)}{\omega_l(m_3^2+m_4^2)}\neq 0$$

It follows from transversality condition that $sign\left(d\,\mathrm{Re}\,\lambda(h_k^j)/dh\right)^{-1}\neq 0$ holds. Further, according to Hopf bifurcation theorem, it follows that the model (3) undergoes Hopf bifurcation around the endemic equilibrium $E^*$ if $\tau$ through to $\tau_l^i$ . So conclusion 2) holds. This completes the proof of theorem 5.

Case(IV) $\tau>0,\ h>0$

Based on the above research methods, two different conditions are discussed.

(1) Let time delay $\tau$ be research parameter, and fix $h^*\in[0,h_1^0)$ , equation (10) becomes

$$\begin{aligned}&\lambda^3+(\mu_0+\xi M^*+\xi_0+2c+\beta I^*+v)\lambda^2+[\mu_0(\xi M^*+\xi_0+2c+\beta I^*+v)\\&+\beta I^*(\xi_0+c+pv)+(c+v)(\xi M^*+\xi_0+c)]\lambda+\mu_0[(c+v)(\xi M^*+\xi_0+\\&c)+\beta I^*(\xi_0+c+pv)]-[\lambda^2+(\xi M^*+\xi_0+c+\mu_0)\lambda+\mu_0(\xi M^*+\xi_0+\\&c)](c+v)e^{-\lambda\tau}+\mu\xi(c+v)I^*e^{-\lambda h^*}=0\end{aligned}\tag{21}$$

Let $\lambda=i\omega\ (\omega>0)$ be the root of equation (21), substituting it into equation (21). And separating real and imaginary part, we have

$$f(\omega)=X^2(\omega)+Y^2(\omega)-m_3^2(\omega)-m_4^2(\omega)=0\tag{22}$$

where

$$\begin{cases} X(\omega) = -\omega^3 + [(v+c+\beta I^*)(\xi_0 + c + \xi M^*) + \mu_0(\xi_0 + c + \xi M^* + \beta I^*)]\omega \\ \qquad - \beta I^*(\xi M^* - pv) - (v+c) I^* \xi \sin \omega h^* \\ Y(\omega) = (\mu_0 + \xi_0 + v + 2c + \xi M^* + \beta I^*)\omega^2 - \mu_0(v + c + \beta I^*)(\xi_0 + c + \xi M^*) \\ \qquad + \beta I^* \mu_0(\xi M^* - pv) - (v+c) I^* \xi \cos \omega h^* \end{cases} \tag{23}$$

$m_3$, $m_4$ are shown in the equation (21).

Supposed that there exists positive roots $\omega_l (l = 0,1,2,...,n)$ of $f(\omega)$, substituting it into equation (23), we have

$$\tau_{1l}^i = \frac{1}{\omega_l} \arccos\left( \frac{m_4 X + m_3 Y}{m_3^2 + m_4^2} \right) + \frac{2i\pi}{\omega_l} \quad (i = 0,1,2,...; l = 0,1,2,...,n)$$

Let $\alpha(\tau_{1l}^i) = 0,\ \omega(\tau_{1l}^i) = \omega_l$, $\tau_{10}^0 = \min\{\tau_{1l}^i\}$.

(2) Let time delay $h$ be research parameter, and fix $\tau^* \in [0, \tau_1^0)$, equation (10) becomes

$$\begin{aligned} &\lambda^3 + (\mu_0 + \xi M^* + \xi_0 + 2c + \beta I^* + v)\lambda^2 + [\mu_0(\xi M^* + \xi_0 + 2c + \beta I^* + v) \\ &+ \beta I^*(\xi_0 + c + pv) + (v+c)(\xi M^* + \xi_0 + c)] + \mu_0[(v+c)(\xi M^* + \xi_0 + \\ &c) + \beta I^*(\xi_0 + c + pv)] - [\lambda^2 + (\mu_0 + \xi M^* + \xi_0 + c)\lambda + \mu_0(\xi M^* + \xi_0 + \\ &c)](v+c)e^{-\lambda\tau^*} + (v+c)\mu\xi I^* e^{-\lambda h} = 0 \end{aligned} \tag{24}$$

Let $\lambda = i\omega (\omega > 0)$ be the root of equation (24), substituting it into equation (24). And separating real and imaginary parts, we have

$$\begin{cases} -\omega^3 + [\mu_0(\xi M^* + \xi_0 + 2c + \beta I^* + v) + \beta I^*(\xi_0 + c + pv) + (v + \\ c)(\xi M^* + \xi_0 + c)]\omega - (v+c)(\xi M^* + \xi_0 + c + \mu_0)\omega \cos \omega\tau^* + \\ [\mu_0(\xi M^* + \xi_0 + c) - \omega^2](v+c)\sin \omega\tau^* = (v+c)\mu\xi I^* \sin \omega h \\ (\mu_0 + \xi M^* + \xi_0 + 2c + \beta I^* + v)\omega^2 - \mu_0[(v+c)(\xi M^* + \xi_0 + \\ c) + \beta I^*(\xi_0 + c + pv)] + (v+c)[\mu_0(\xi M^* + \xi_0 + c) - \omega^2)]\cos \omega\tau^* \\ + (v+c)(M^* + \xi_0 + c + \mu_0)\omega \sin \omega\tau^* = (v+c)\mu\xi I^* \cos \omega h \end{cases} \tag{25}$$

And removing $\sin \omega h$ and $\cos \omega h$, we obtain

$$f(\omega) = X_1^2(\omega) + Y_1^2(\omega) - (v+c)^2 \mu^2 \xi^2 I^{*2} \tag{26}$$

where

$$\begin{aligned} X_1(\omega) = &-\omega^3 + [\mu_0(\xi M^* + \xi_0 + 2c + \beta I^* + v) + \beta I^*(\xi_0 + c + pv) + (\xi M^* \\ &+ \xi_0 + c)(v+c)]\omega - (v+c)(\xi M^* + \xi_0 + c + \mu_0)\omega \cos \omega\tau^* + (v + \\ &c)[\mu_0(\xi M^* + \xi_0 + c) - \omega^2]\sin \omega\tau^* \end{aligned}$$

$$Y_1(\omega) = (\mu_0 + \xi M^* + \xi_0 + 2c + \beta I^* + v)\omega^2 - \mu_0[(\xi M^* + \xi_0 + c)(v+c)$$
$$+ \beta I^*(\xi_0 + c + pv)] + (v+c)[\mu_0(\xi M^* + \xi_0 + c) - \omega^2)]\cos\omega\tau^*$$
$$+ (v+c)(M^* + \xi_0 + c + \mu_0)\omega\sin\omega\tau^*$$

Supposed that there exists positive roots $\omega_k (k = 0,1,2,...,m)$ of $f(\omega)$, substituting it into equation (23), we have

$$h_{1k}^j = \frac{1}{\omega_k}\arccos\left(\frac{Y_1}{(v+c)\mu\xi I^*}\right) + \frac{2j\pi}{\omega_k}, \quad (j = 0,1,2,...; k = 0,1,2,...,m)$$

Let $\alpha(h_{1k}^j) = 0,\ \omega(h_{1k}^j) = \omega_k$ and $h_{10}^0 = \min\{h_{1k}^j\}$.

In similar to the proof of theorem 4, we obtain

**Theorem 6** If $R_0 > 1,\ \tau > 0,\ h > 0$ holds, we have

1) If $h^* \in [0, h_1^0)$, and equation (22) exists positive roots, when $\tau \in [0, \tau_{10}^0)$, the positive equilibrium point $E^*$ of model (3) is asymptotically stable; when $\tau > \tau_{10}^0$, the equation (23) has at least a positive real part characteristic root, then the positive equilibrium point $E^*$ of model (3) is unstable; when $\tau = \tau_{1l}^i\ (l = 0,1,2,...,n;\ i \in N)$, then the model (3) undergoes Hopf bifurcation around the positive equilibrium point $E^*$.

2) If $\tau^* \in [0, \tau_1^0)$, and equation (26) exists positive roots, when $h \in [0, h_{10}^0)$, the positive equilibrium point $E^*$ of model (3) is asymptotically stable; when $h > h_{10}^0$, the equation (24) has at least a positive real part characteristic root, then the positive equilibrium point $E^*$ of model (3) is unstable; when $h = h_{1k}^j,\ k = 0,1,2,...,m;\ j \in N$, then there exists Hopf bifurcation around the endemic equilibrium $E^*$ of the model (3).

# 4. Numerical simulations

To check the feasibility of our analysis, we present some numerical computations in the section by using Matlab. To do numerical simulation of the disease-free equilibrium $E_0$, we choose the parameter values as follows: $c = 0.007$, $\beta = 0.2$, $\xi = 0.6$, $\xi_0 = 0.3$, $p = 0.6$, $v = 0.29$, $\mu = 0.2$, $\mu_0 = 0.6$, and let $(0.4, 0.3, 0)$ be initial value, we can obtain $R_0 < 1$. According to Theorem 2, it follows that the disease-free equilibrium $E_0$ of model (3) is globally asymptotically stable, and it has nothing to do with the value of $\tau$ and $h$. It means that the disease disappears (see Fig. 1).

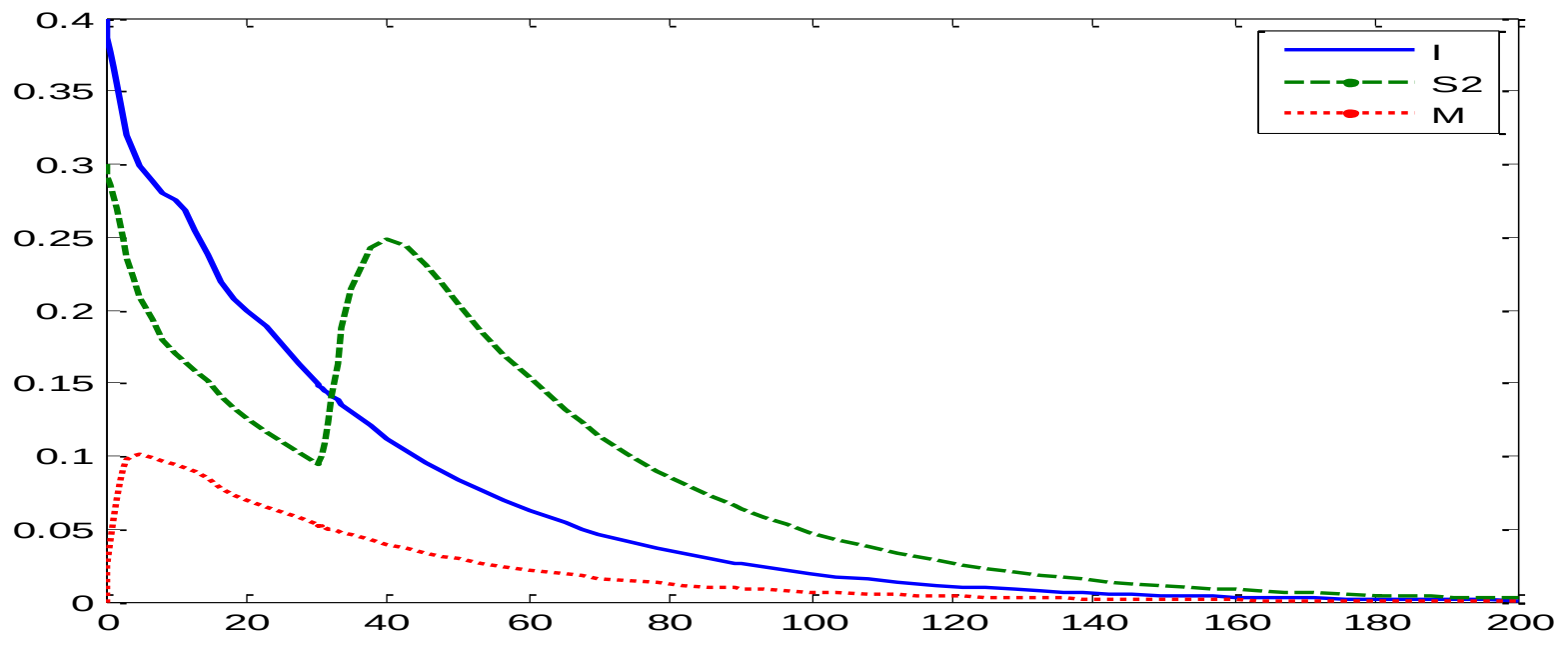

Fig.1 The globally stable mimetic diagram of the disease-free equilibrium $E_0$

We present numerical simulation for the endemic equilibrium $E^*$ by choosing the following set of parameter values $c = 0.005$, $\beta = 0.5$, $\xi = 0.5$, $\xi_0 = 0.02$, $p = 0.4$, $v = 0.2$, $\mu = 0.2$, $\mu_0 = 0.02$. According to the parameter selection, respectively discuss the endemic equilibrium. By calculations, we have $R_0 > 1$, then the unique endemic equilibrium $E^*(0.013, 0.487, 0.13)$ of model (3) exists.

Case(I) $\tau = h = 0$

According to the condition of Theorem 3, it follows that the endemic equilibrium $E^*$ of model (3) is stable(see Fig. 2).

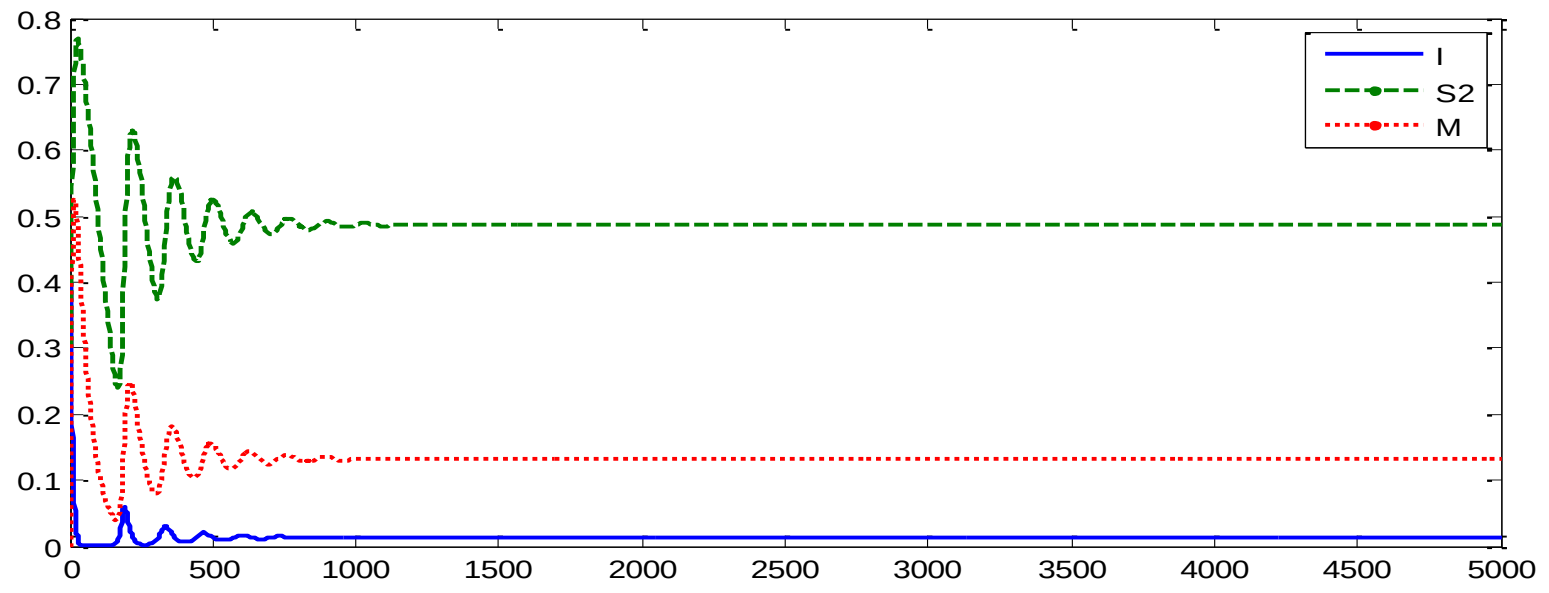

Fig.2 The stable mimetic diagram of the endemic equilibrium $E_0$

Case(II) $\tau = 0$, $h > 0$

By calculations, we obtain $\pm i\omega_0 \approx \pm 0.045i$ and $h_1^0 \approx 146.88$. And we select $h = 100 < h_1^0$, $h = 372 > h_1^0$, its mimetic diagrams are illustrated in Fig. 3(1)-3(3). With the increase of $h$, the endemic equilibrium $E^*$ of model (3) from stable state to unstable state by damped oscillation.

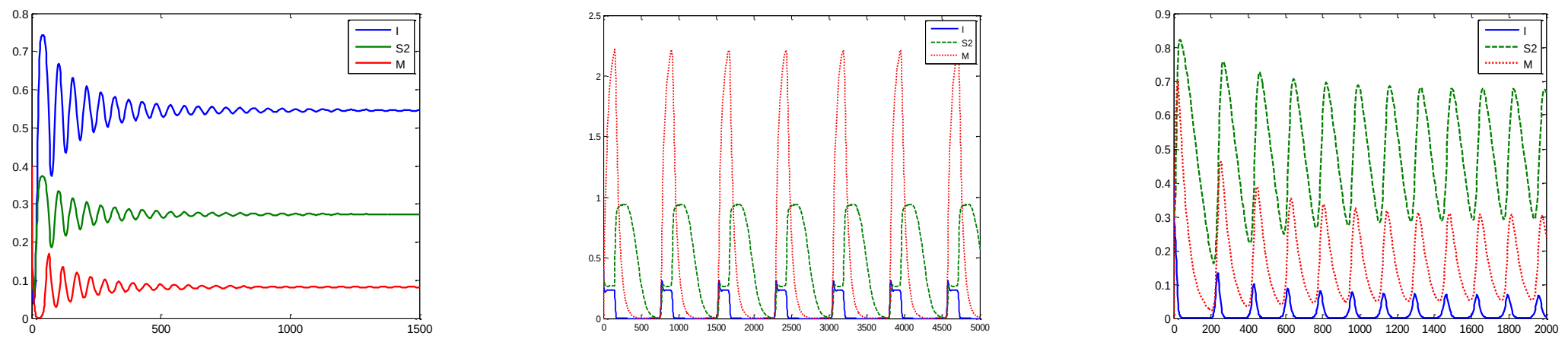

Fig.3(1)When $h = 100$, $E^*$ is globally stable.

Fig.3(2)When $h = 146.88$, periodic solution exists.

Fig.3(3)When $h = 372$, $E^*$ is unstable.

Case(III) $\tau > 0,\ h = 0$

We obtain $\pm i\omega_0 \approx \pm 0.3627i$ and $\tau_1^0 \approx 66.37$ . And we select $\tau = 30 < \tau_1^0$ , $\tau = 110.32 > \tau_1^0$ , its mimetic diagrams are illustrated in Fig. 4(1)-4(3). With the increase of $h$, the endemic equilibrium $E^*$ of model (3) from stable state to unstable state by damped oscillation.

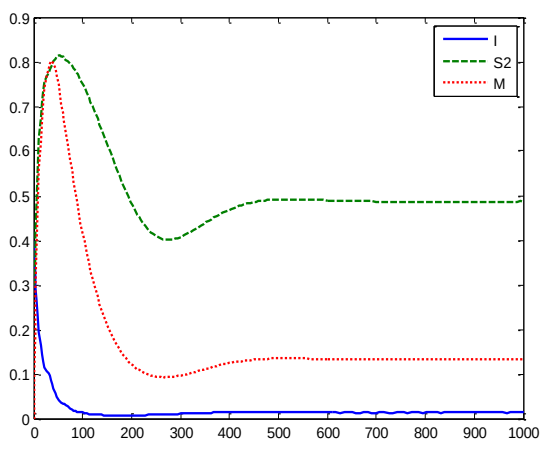

Fig.4(1)When $\tau = 30$, $E^*$ is globally stable.

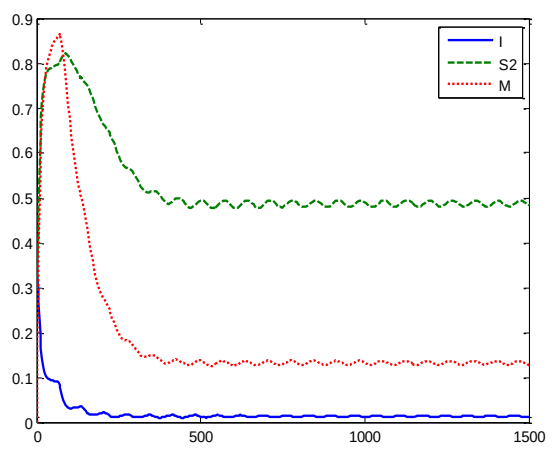

Fig.4(2)When $\tau = 66.37$, periodic solution exists.

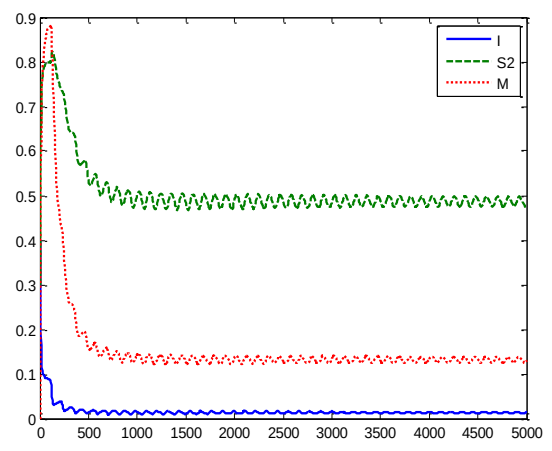

Fig.4(3)When $\tau = 110.32$, $E^*$ is unstable.

Case(IV) $\tau > 0,\ h > 0$

Firstly, let $h = 100$ ,we calculate $\tau_{10}^0 \approx 74.35$ , And we select $\tau = 50 < \tau_{10}^0$ , $\tau = 80, 110, 200 > \tau_{10}^0$. The trajectory of $E^*$ see Fig.5(1)-5(5).

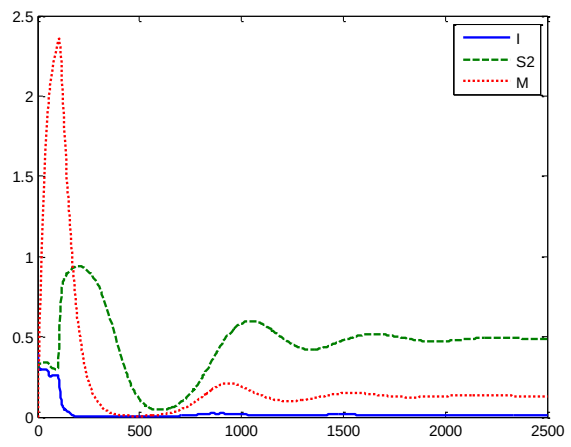

Fig.5(1)When $\tau = 50$, $E^*$ is globally stable.

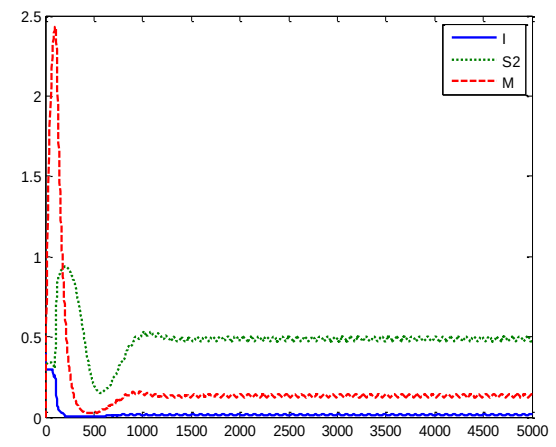

Fig.5(2)When $\tau = 74.35$, periodic solution exists.

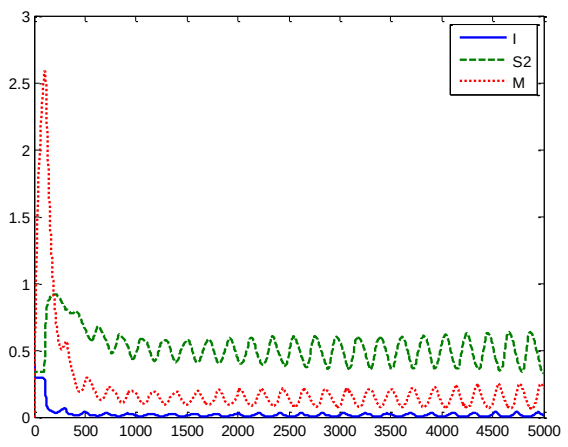

Fig.5(3)When $\tau = 80$, $E^*$ is unstable.

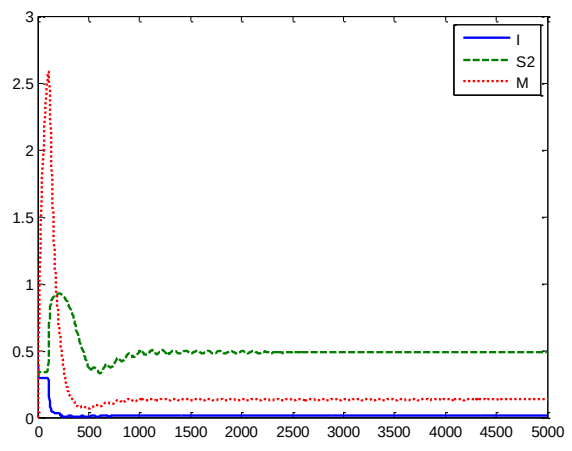

Fig.5(4)When $\tau = 110$, $E^*$ is globally stable.

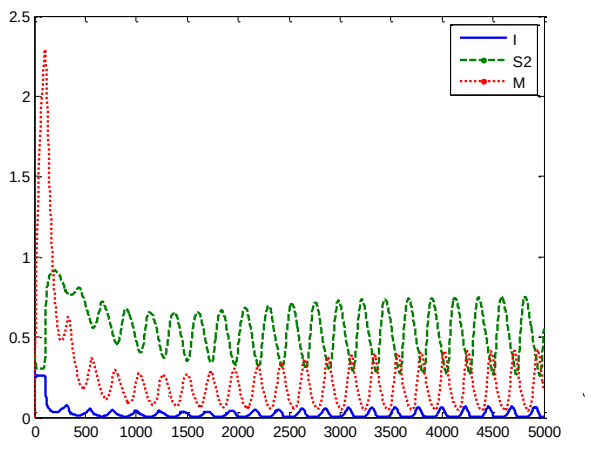

Fig.5(5)When $\tau = 200$, $E^*$ is unstable.

With the increase of $\tau$, $E^*$ transfers from stable tend to equilibrium point $E^*$ by damped oscillation (see Fig.5(1)-5(3)). To increase the value of $\tau$, the curve tend to be stable. Compared with Fig.5(1), the time of tending to $E^*$ is shorted(see Fig. 5(4)). To increase the value of $\tau$ again, the curve appears periodic oscillation(see Fig. 5(5)). Based on a given parameter $h$, it can be seen that delay time $\tau$ make the endemic equilibrium $E^*$ from stable to unstable or from unstable to stable. As a result, the time delay $\tau$ plays a role in adjusting the stability of endemic equilibrium $E^*$.

Let $\tau = 110$, we calculate $h_{10}^0 \approx 122.58$, And we select $h = 100 < h_{10}^0$, $h = 240 > h_{10}^0$, see Fig.6(1)-6(3).

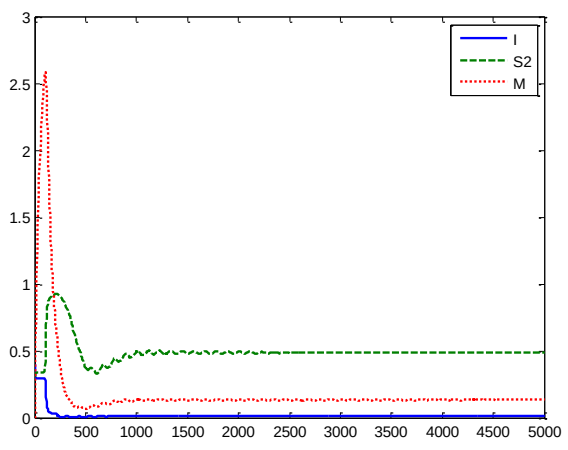

Fig.6(1)When $h = 100$, $E^*$ is globally stable.

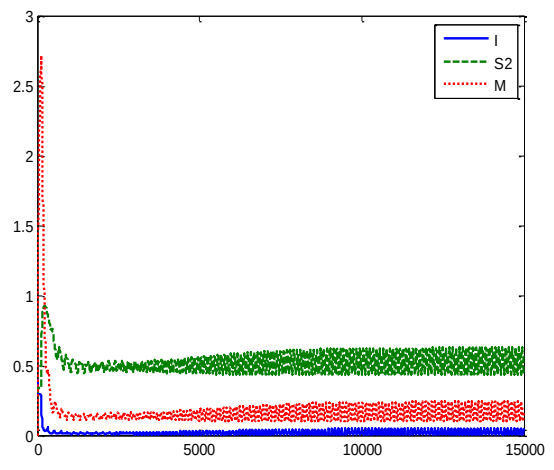

Fig.6(2)When $h = 122.58$, periodic solution exists.

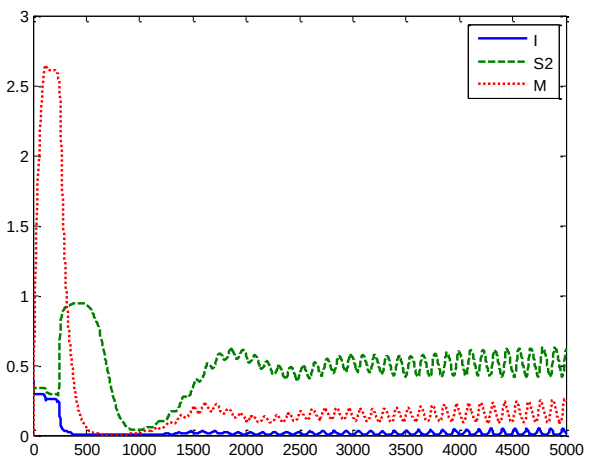

Fig.6(3)When $h = 240$, $E^*$ is unstable.

From Fig. 6(1)-6(3), we obtain $E^*$ from stable tend to equilibrium point $E^*$ by damped oscillation with the increase of $\tau$. To increase the value of $\tau$ again, $E^*$ becomes unstable.

# 5. The Application of the Model of in influenza A H1N1 virus

## 5.1 The statement of problem

Influenza A H1N1 virus includes gene fragments of three kinds of influenza virus of human, avian, swine flu, it spreads mainly through body fluids, droplets and air, latent period is generally 1 to 7 days[15]. With the spread of the disease, the number of patients also increases, it brings significant impact on the quality of life and the development of social economy. In the course of the global fight against H1N1, there are a lot of research about the dissemination and the development trend of the disease. Lin etc. analyze the trend of international flu in statistical method, and assess the impact of pandemic on the country, but they do not come up with effective strategies to control the disease[16]. Based on the research results of infectious disease model, this paper quantitatively analyzes the influence of the parameters such as the rate of consciousness transfer and the implementation rate of the media project on the development trend of the disease, and puts forward the prevention and control strategies.

## 5.2 The determine of parameters

Based on data from the China Statistical Yearbook "birth rate, death rate and natural growth rate" in 2012, we know that the rate of constant input and natural death is $c = 0.007$[17] .The value of $v$ references [18], for other parameter values ,we let $p = 0.6$, $\xi_0 = 0.5$, $\mu_0 = 0.7$, $\beta = 0.3$. The awareness transfer rate $\xi$, the media project implementation rate $\xi_0$, the time delay $\tau$ and $h$ are mainly considered for controlling infectious diseases.

5.3 Numerical simulations

With the change of awareness transfer rate, media project implementation rate, latent period and the time delay of media, the development trend of disease will change. The trend of disease is analyzes in different parameter values by numerical simulations. The parameter values are shown Table 1.

Table 1 The influence of parameter $\xi$, $\mu$, $\tau$, $h$ on disease

| $\xi$ | $\mu$ | $\tau$ | $h$ | Simulation diagram | The change of infected individuals |
|---|---|---|---|---|---|
| 0.5 | 0.8 | 7 | 35 | Fig.7(1) | The time lag of media is longer, the elimination of infected individuals becomes longer |
| | | | 20 | Fig.7(2) | |
| 0.01 | 0.8 | 2 | 20 | Fig.7(3) | The latent period is longer, the elimination of infected individuals becomes longer |
| 0.5 | | | | Fig.7(4) | The transfer rate decreases, the elimination of infected individuals becomes slower |
| 0.5 | 1 | 2 | 20 | Fig.7(5) | The media project implementation rate increases, the elimination of infected individuals becomes faster |

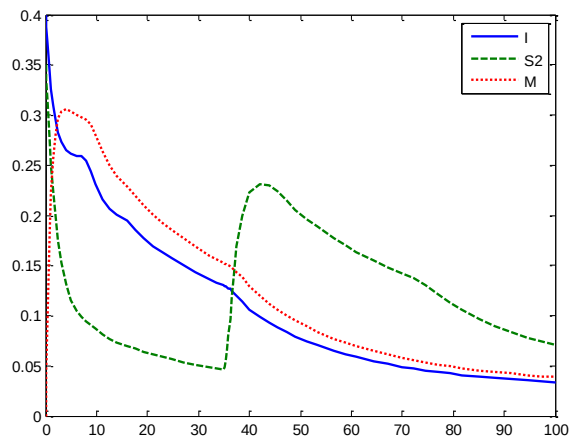

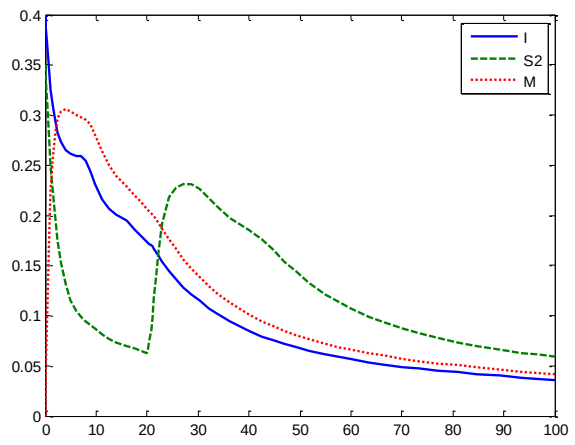

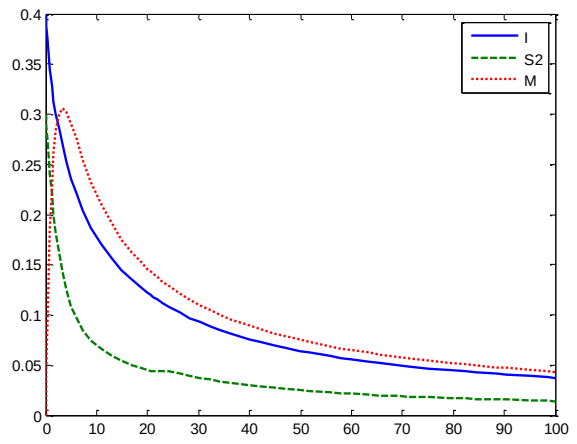

Fig.7(1) Fig.7(2)

Fig.7(3)

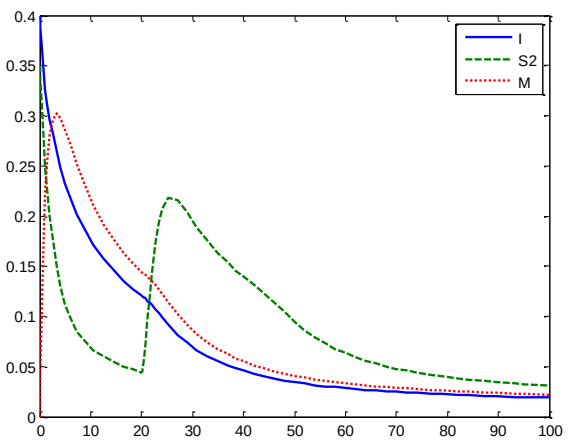

Fig.7(4)

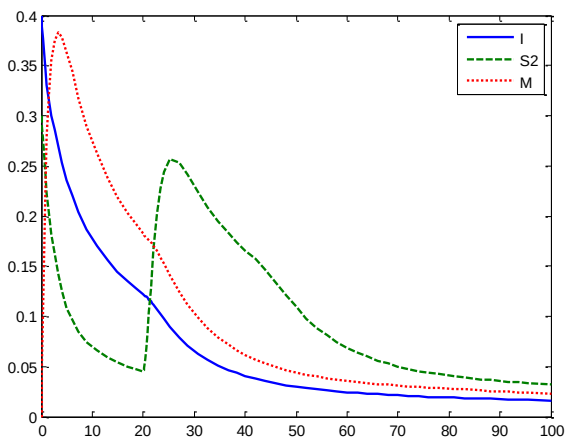

Fig.7(5)

Fig.7(1) -7(5) The variation diagram of infected individuals

The results of numerical simulations show that we must increase media propaganda after the disease outbreaks, let the public know the way of disease transmission route and make prevention measures as soon as possible, so it may effectively control the infectious disease.

5.4 The strategies of control disease

For different diseases, the comprehensive control strategies are implemented, it reduces disease incidence.

(1) After the outbreak of the disease, the processing time of disease information is not too long, the disease is reported to raise the consciousness of prevention as soon as possible. It makes that the relevant departments of medical takes timely measures to control the disease, and the time of disease stabilization is shortened;

(2) The media propaganda increases, and the rate of consciousness transfer and the implementation rate of the media project also increase. It makes the proportion of infected individuals reduce significantly and the disease is controlled effectively.

(3) Because there exists a latent period for the disease, it is difficult to control the disease. So people should try to avoid going to public places and reduce the chance of being infected;

(4) It can effectively prevent the spread of the disease by isolating infection individuals. And it can improve the cure rate of the patients by processing medical assistance and increasing the vaccination rate.